\documentclass[fleqn,12pt,twoside]{article}
\usepackage{espcrc1}
\usepackage{epsf}

\usepackage{graphicx}
\usepackage[figuresright]{rotating}

\newcommand{\be}{\begin{equation}}

\renewcommand{\vec}[1]{{\bf #1}}
\newcommand{\ee}{\end{equation}}
\newcommand{\bea}{\begin{eqnarray}}
\newcommand{\eea}{\end{eqnarray}}

\newcommand{\vp}{{\bf p}}

\title{Elliptic flow and freeze-out from the parton cascade MPC}

\author{D\'enes Moln\'ar and Miklos Gyulassy\address[CU]{Physics Department, Columbia University\\538 West 120th Street, New York, NY 10027, U.S.A.}%
  \thanks{This work was supported by the Director, Office of Energy Research,
          Division of Nuclear Physics of the Office of High Energy and Nuclear
          Physics of the U.S. Department of Energy under contract No.
          DE-FG-02-93ER-40764.
  }
}

\begin{document}

\maketitle

\begin{abstract}
\small
Differential elliptic flow out to $p_\perp \sim 5$ GeV/c
and particle spectra are calculated using the MPC elastic parton cascade
model for Au+Au at $E_{cm}\sim 130$ $A$ GeV.
The evolution is computed from parton transport theory,
followed by hadronization either via independent fragmentation or
by imposing parton-hadron duality.
With pQCD elastic cross sections,
very large initial gluon densities $dN_g/d\eta > 7000$
are required to reproduce the data measured by the STAR
collaboration.
In addition, elliptic flow and the $p_\perp$ spectra are shown to be very sensitive
to particle subdivision.
\end{abstract}

\section{Introduction}
\label{Section:intro}

Elliptic flow, $v_2(p_\perp)=\langle \cos(2\phi)\rangle_{p_\perp}$,
 the differential second moment of the azimuthal momentum distribution,
has been the subject of increasing interest
\cite{hydro,Zhang:1999rs,molnar_v2,gvw,v2_cascade},
especially
since the discovery\cite{STARv2} at RHIC that $v_2(p_\perp>1 $ GeV)
 $\rightarrow 0.2$.
This sizable  high-$p_\perp$ collective effect
 depends strongly on the dynamics in a heavy
ion collision and provides important information about the 
density and effective energy loss of partons.

The simplest theoretical framework to study elliptic flow is ideal
hydrodynamics\cite{hydro}.
For RHIC energies, ideal hydrodynamics
agrees remarkably well with the measured elliptic flow data%
\cite{STARv2}
up to transverse momenta $\sim 1.5$
GeV$/c$.
However,
it fails to saturate at high $p_\perp>2$ GeV as does the data reported
by STAR at Quark Matter 2001.

A theoretical problem with ideal hydrodynamics is that it
assumes local equilibrium throughout the whole evolution.
This idealization is marginal
for conditions encountered in heavy ion collisions\cite{nonequil}.
A theoretical framework is required that allows for nonequilibrium dynamics.
Covariant  Boltzmann transport theory provides a convenient framework
that depends on the local mean free path $\lambda(x) \equiv 1/\sigma n(x)$.

Parton cascade simulations\cite{Zhang:1999rs,molnar_v2}
show on the other hand, that  the initial parton density
based on HIJING\cite{Gyulassy:1994ew}
is too low  to produce the observed  elliptic flow
unless the pQCD cross sections are artificially enhanced 
by a factor $\sim 2-3$.
However,
gluon saturation models\cite{Eskola:2000fc}
predict up to five times higher initial densities,
and these  may be dense enough to generate the observed collective flow
even with pQCD elastic cross sections.
In this study,
we explore the dependence of elliptic flow on the initial density
and the elastic $gg$ cross section.

Calculations
based on inelastic parton energy loss\cite{gvw}
also predict saturation or  decreasing $v_2$ at high $p_\perp$.
These calculations are only valid for high $p_\perp$,
where collective transverse flow from lower-$p_\perp$ partons can be neglected.
The collective component from low $p_\perp$
is, 
on the other hand,
automatically incorporated in parton cascades.
Though parton cascades
lack at present covariant
inelastic energy loss, 
elastic energy loss alone may  account
for the observed high-$p_\perp$ azimuthal flow pattern
as long as the number of elastic collisions is 
large enough \cite{molnar_v2}.

Forerunners of this study
\cite{Zhang:1999rs,molnar_v2}
computed elliptic flow
for partonic systems starting from initial conditions expected at RHIC.
Here, we extend these in three aspects.
We compute the $p_\perp$-differential elliptic flow  $v_2(p_\perp)$,
model hadronization,
which enables us to compute the final observable hadron flow,
and consider realistic nuclear geometry.

\section{Covariant parton transport theory}
\label{Section:transport_theory}

We consider here, as in \cite{molnar_v2,nonequil,Yang,Zhang:1998ej},
the simplest form of Lorentz-covariant Boltzmann transport theory
in which the on-shell phase space density $f(x,\vp)$,
evolves with an elastic $2\to 2$ rate.
We solve the transport equation using the cascade method,
which inherently violates Lorentz covariance.
To ensure Lorentz covariance, we apply the parton subdivision
technique\cite{Yang,Zhang:1998tj}.
See \cite{nonequil} for details.

The elastic gluon scattering matrix elements in dense parton systems
are modeled by a Debye-screened form
$
d\sigma/dt
  = \sigma_0
    (1 + \mu^2/s)
    \mu^2/(t-\mu^2)^2$,
where $\mu$ is the screening mass,
$\sigma_0 = 9\pi\alpha_s^2/2\mu^2$ is the total cross section,
which we chose to be independent of energy.

\section{Numerical results}
\label{Section:glue_results}

The initial condition was a longitudinally boost invariant Bjorken tube
in local thermal equilibrium
at $T_0=700$ MeV at proper time $\tau_0=0.1$~fm/$c$
as by fitting the gluon mini-jet transverse momentum spectrum
predicted by HIJING\cite{Gyulassy:1994ew}
(without shadowing and jet quenching). 
The  pseudo-rapidity distribution
was taken as uniform between $|\eta| < 5$,
while
the transverse density distribution was assumed to be proportional
to the binary collision
distribution for two Woods-Saxon distributions.
For collisions at impact parameter $b$,
the  transverse binary collision profile is  
$
dN_g({\bf b})/d\eta d^2 {\bf x}_\perp = \sigma_{jet}
T_A({\bf x}_\perp + {\bf b}/2)T_A({\bf x}_\perp - {\bf b}/2)$,
where $T_A({\bf b})=\int dz\; \rho_A(\sqrt{z^2+{\bf b}^2})$,
in terms of the diffuse nuclear density $\rho_A(r)$. 
From  HIJING, the pQCD jet cross section
normalization at $b=0$,
$\sigma_{jet}$ is fixed such that  $dN(0)/d\eta=210$.

The evolution was performed numerically 
with 40 and 100 mb isotropic cross sections,
and with 3, 40 and 100 mb gluonic cross sections ($\mu=T_0$).
We used particle subdivision
$\ell=100$ for impact parameters 0, 2, and 4 fm,
while $\ell=220$, 450, 1100, and 5000,
for $b=6$, 8, 10, and 12 fm.

Two different hadronization schemes were applied.
One is based on local parton-hadron duality,
where as \cite{Eskola:2000fc},
we assumed that each gluon gets converted
to a pion with equal probability for the three isospin states.
Hence,
$
f_{h^-}(\vec p_\perp) \approx f_{\pi^-}(\vec p_\perp) = \frac{1}{3} f_{g}(\vec p_\perp).
$

The other hadronization prescription was independent fragmentation.
We considered only the $g\to \pi^{\pm}$ channel
with the NLO
fragmentation function from \cite{Binnewies:1995ju}.
For the scale factor $s \equiv \log(Q^2)/\log(Q_0^2)$ we take $s=0$
because  the initial HIJING gluon distribution is already quenched
due to initial and final state radiation.
Since we do not consider soft physics, we limit our study to hadrons
with $p_\perp > 2$ GeV.

Due to scaling \cite{nonequil},
differential elliptic flow depends on $\sigma_0$ and $dN_g/d\eta$ only via
$\xi\equiv\sigma_0 dN_g(0)/d\eta$.
On the other hand,
the $p_\perp$ spectrum depends on
$\sigma_0$ and $dN_g/d\eta$ {\em separately}.

\subsection{Elliptic flow results}

\begin{figure}[h]
\center
\leavevmode
\hbox{
    \epsfysize 4.5cm
    \epsfbox{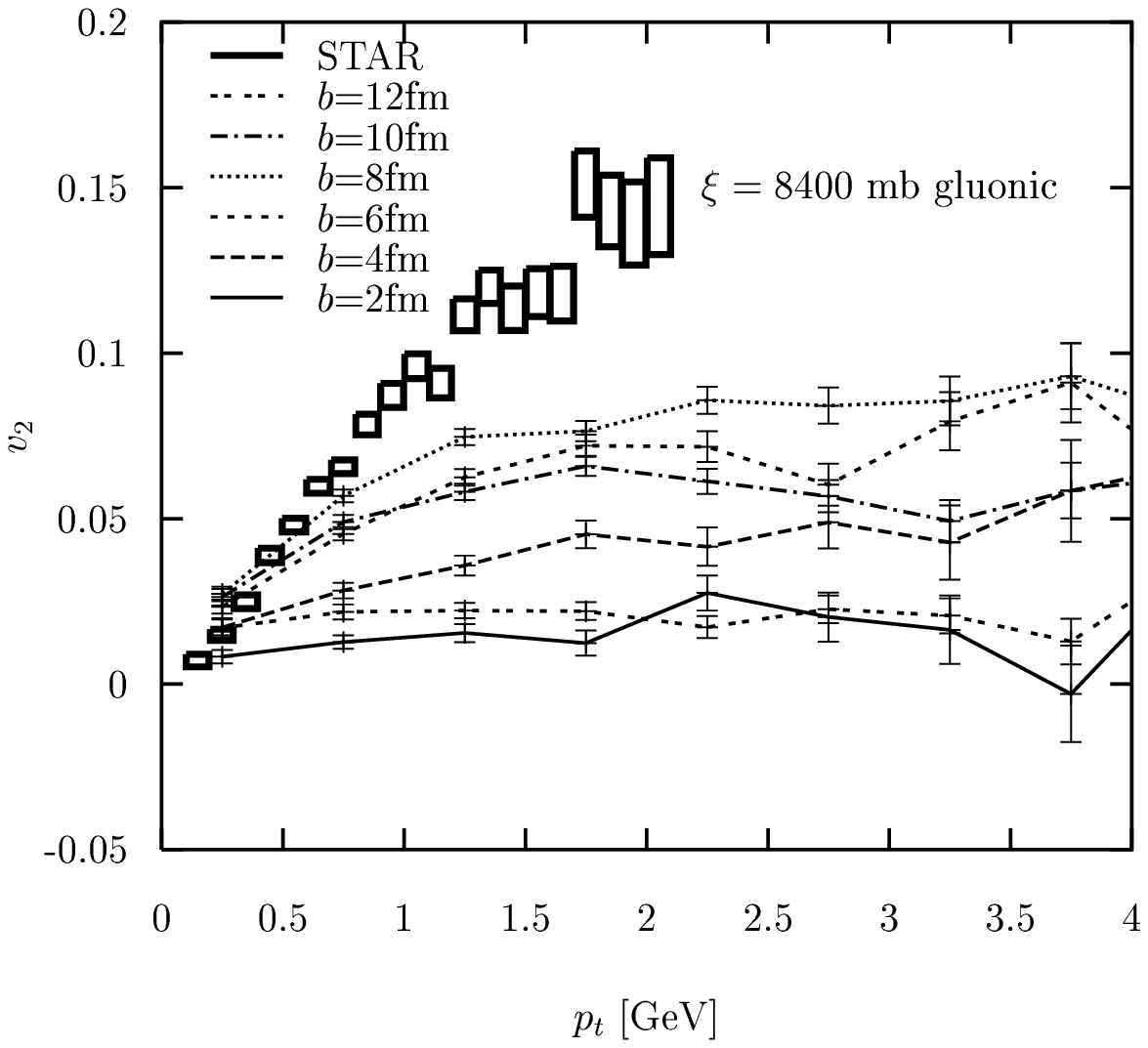}
    \epsfysize 4.5cm
    \epsfbox{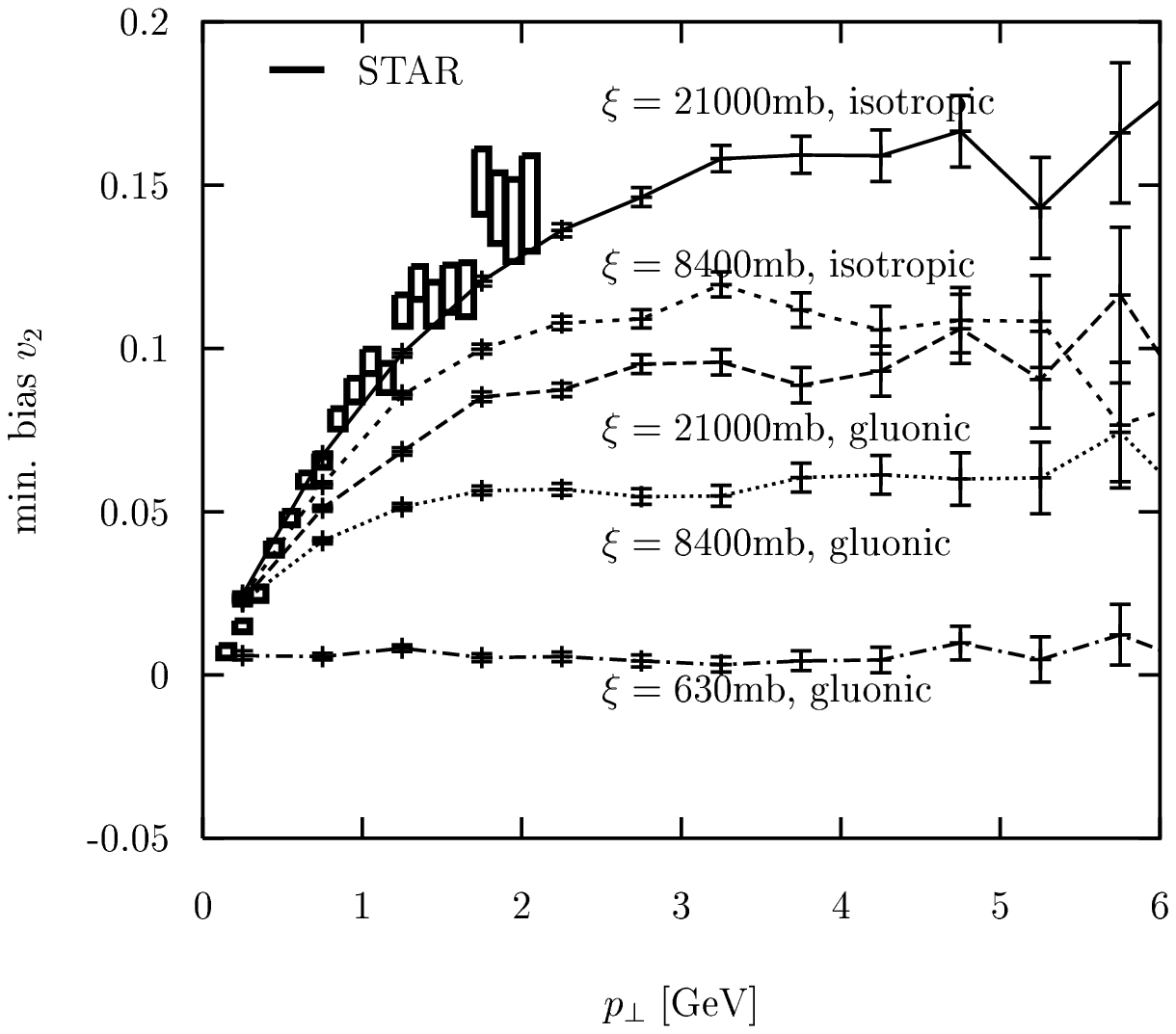}
    \epsfysize 4.5cm
    \epsfbox{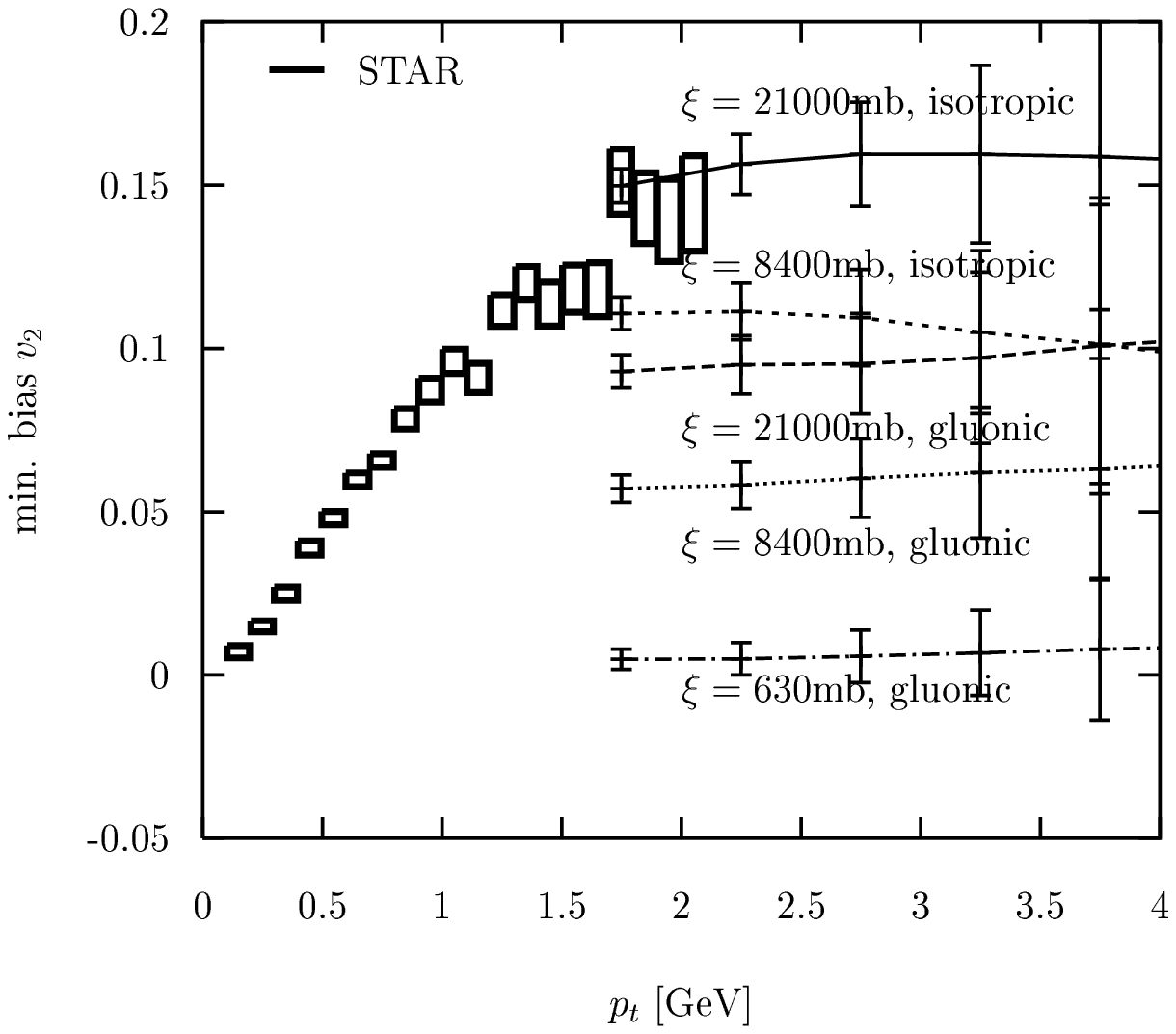}
}
\vspace*{-1.5cm}
\caption{
\footnotesize
Elliptic flow as a function of $p_\perp$ for Au+Au
at $\sqrt{s}=130A$ GeV.
Left figure shows gluon elliptic flow for different impact parameters.
Middle and right figures show {\em minimum-bias} charged hadron $v_2$ 
for hadronization
via local parton-hadron duality (middle) or
independent fragmentation (right).
}
\label{Figure:1}
\label{Figure:v2}
\end{figure}

Fig. \ref{Figure:v2} shows elliptic flow as a function of $p_\perp$ and
impact parameter.
With increasing $p_\perp$,
the minimum-bias elliptic flow increases until $p_\perp\sim 1.5-2$ GeV,
where it saturates.
With increasing impact parameter,
elliptic flow first monotonically increases,
then monotonically decreases,
showing a maximum at $b\approx 8$ fm.

\vspace*{-0.5cm}
\vspace*{-0.7cm}

The minimum-bias flow was computed via
$
v_2^{m.b.}(p_\perp) \equiv \frac{2\pi}{\pi b_{max}^2} \int_0^{b_{max}} v_2(b,p_\perp)\; b\; db,
$
where we chose $b_{max} = 12$ fm.
This differs from the definition by STAR,
which weights $v_2(b,p_\perp)$ with $dN_g/dydp_\perp$
and hence results in a smaller $v_2(p_\perp)$ than our estimate.

The $\xi=8400$ mb isotropic curve is almost the same as the $\xi=21000$ mb
gluonic curve, showing that not the {\em total}, rather
the {\em transport} cross section is the relevant parameter.
Thus,
we expect that the flow in the $\xi=21000$ mb isotropic case 
can also reproduced with $\xi\sim 50000$ mb and a gluonic cross section.
With the pQCD $gg$ cross section of 3mb,
this corresponds to an 80 times larger initial density than from HIJING.

\subsection{Particle spectra}
\label{Subsection:glue_spectra}

\begin{figure}[h]
\center
\leavevmode
\hbox{
    \epsfysize 5.5cm
    \epsfbox{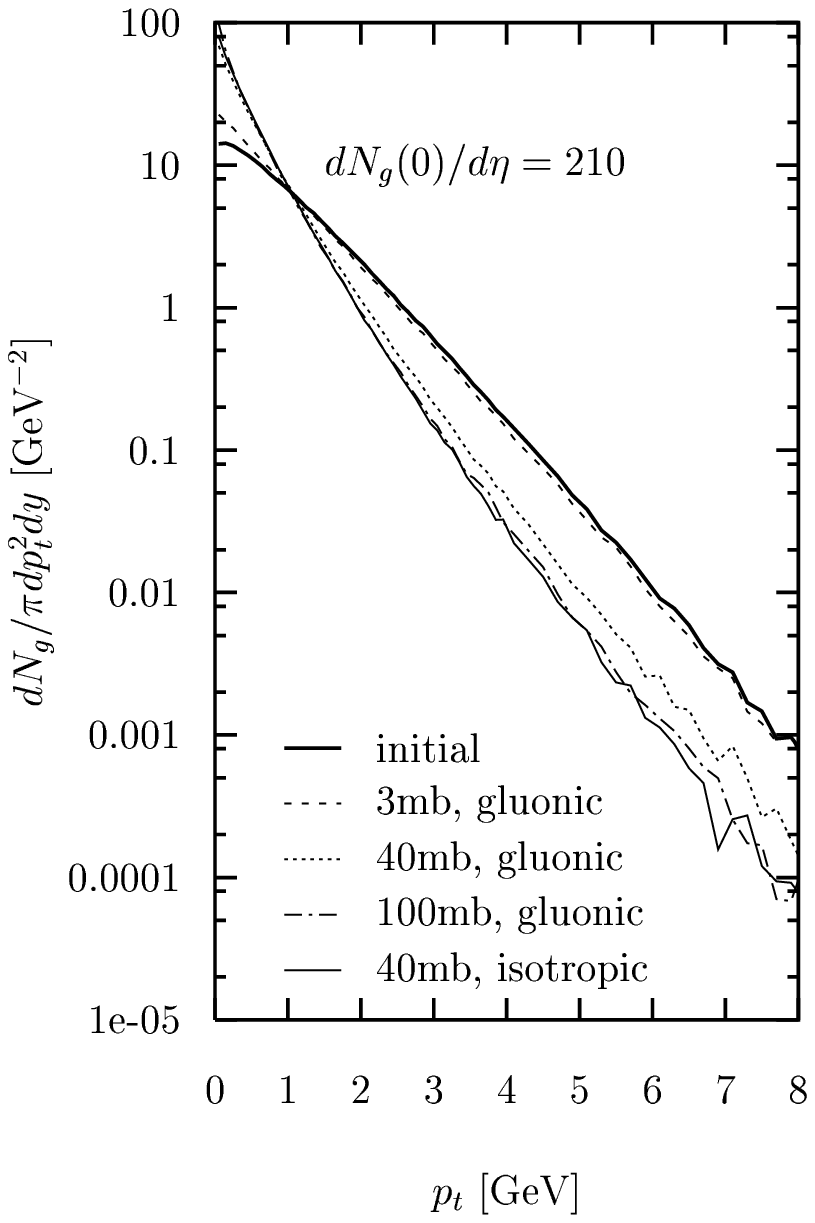}
    \epsfysize 5.5cm
    \epsfbox{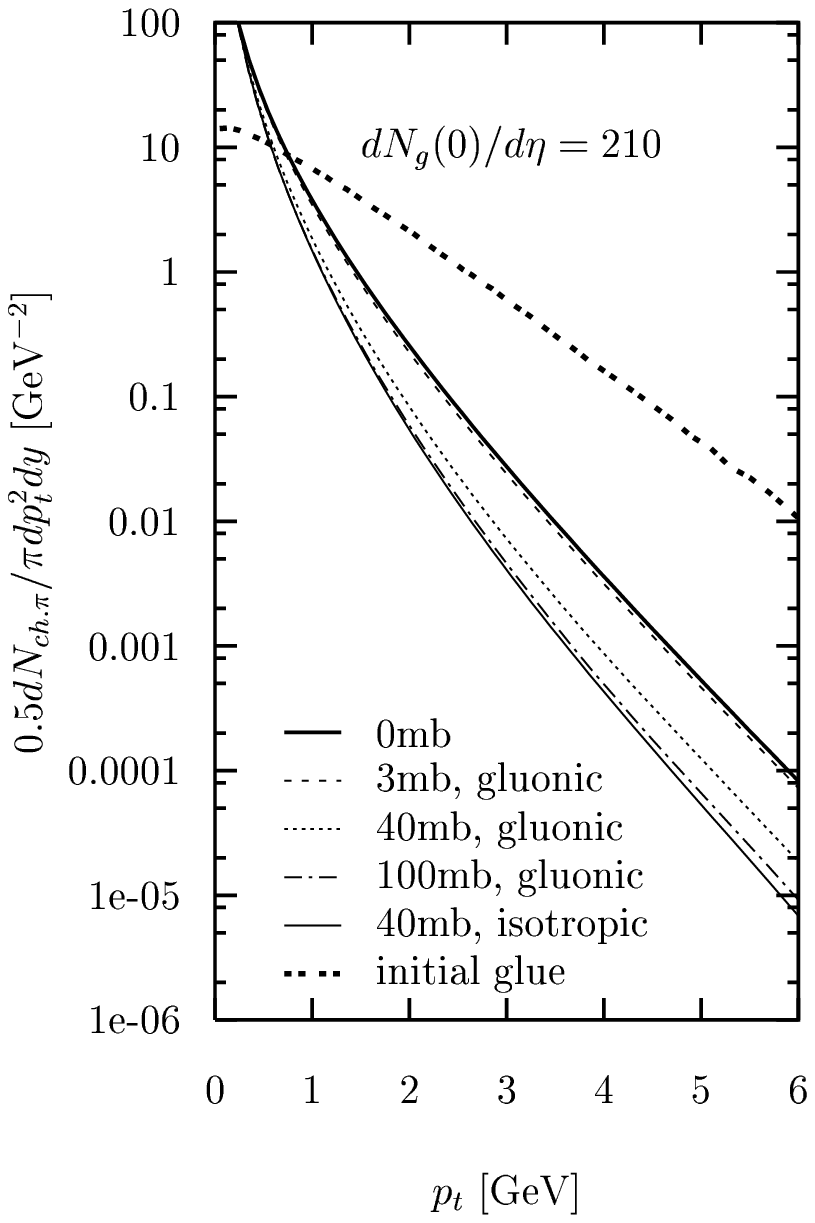}
    \epsfysize 5.5cm
    \epsfbox{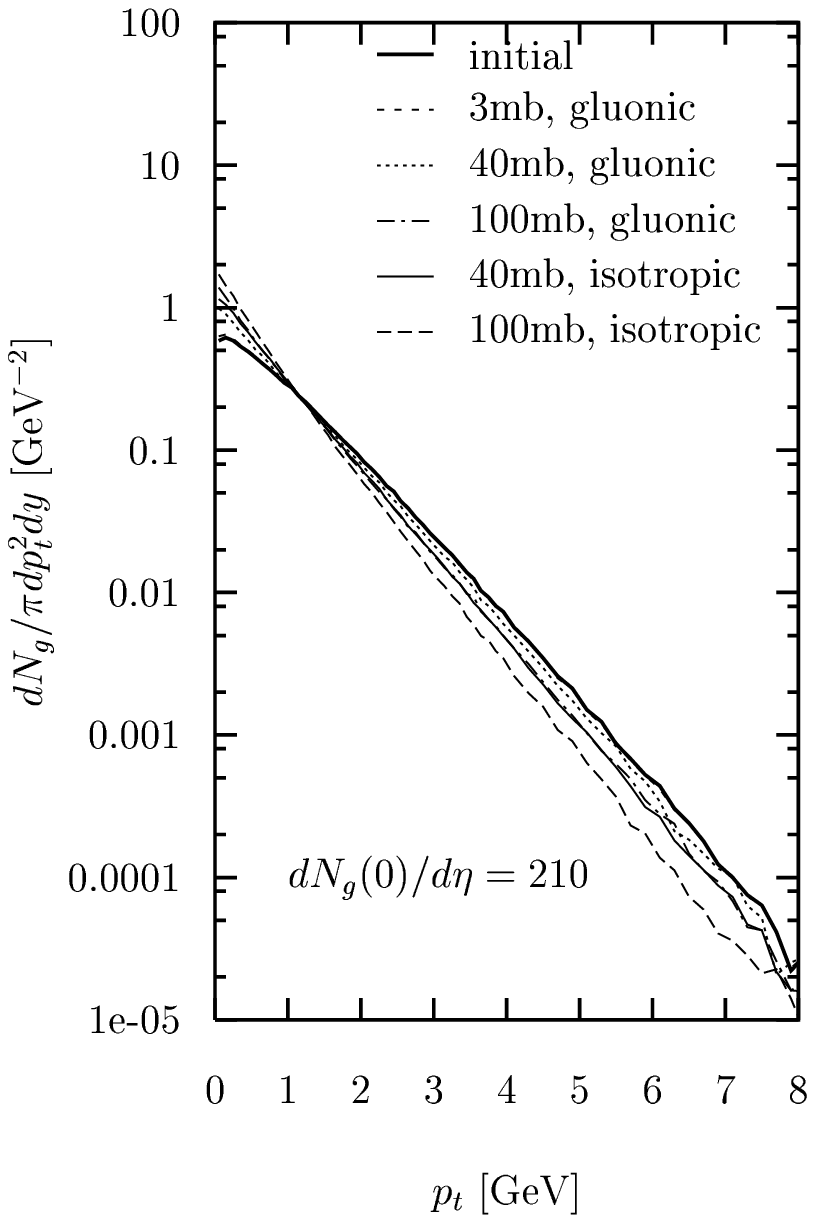}
    \epsfysize 5.5cm
    \epsfbox{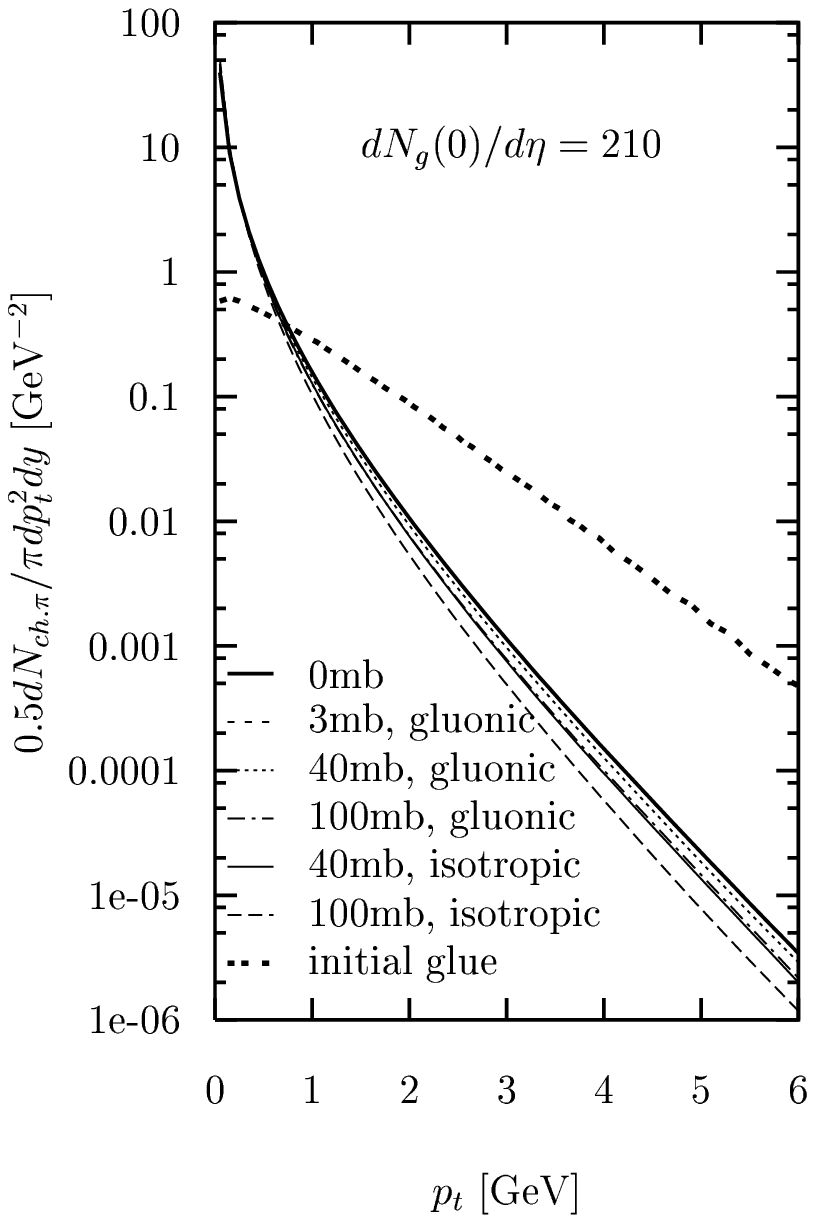}
}
\vspace*{-1.5cm}
\caption{
\footnotesize
Final gluon and negative hadron $p_\perp$ spectra (via indep. fragmentation)
as a function of the elastic cross section
for Au+Au at $\sqrt{s} = 130A$ GeV
with $b=6$ fm (left two plots) and $b=12$ fm (right two plots).
For local parton-hadron duality, hadron spectra are proportional to gluon spectra.
\vspace*{-0.5cm}
}
\label{Figure:2}
\label{Figure:pt}
\end{figure}

Fig. \ref{Figure:pt} shows
the quenching of the $p_\perp$ spectra 
due to elastic energy loss.
In the dilute $\xi =630$ mb case quenching is negligible,
however, for $\xi = 21000$ mb it is 
an order of magnitude at $p_\perp > 6$ GeV for semicentral collisions.
Naturally, for a fixed $\xi$, quenching decreases 
as the collision becomes more peripheral
because the parton density decreases.
Hadronization via independent fragmentation results
in an additional quenching.

\begin{figure}[h]
\center
\leavevmode
\hbox{
    \epsfysize 4.5cm
    \epsfbox{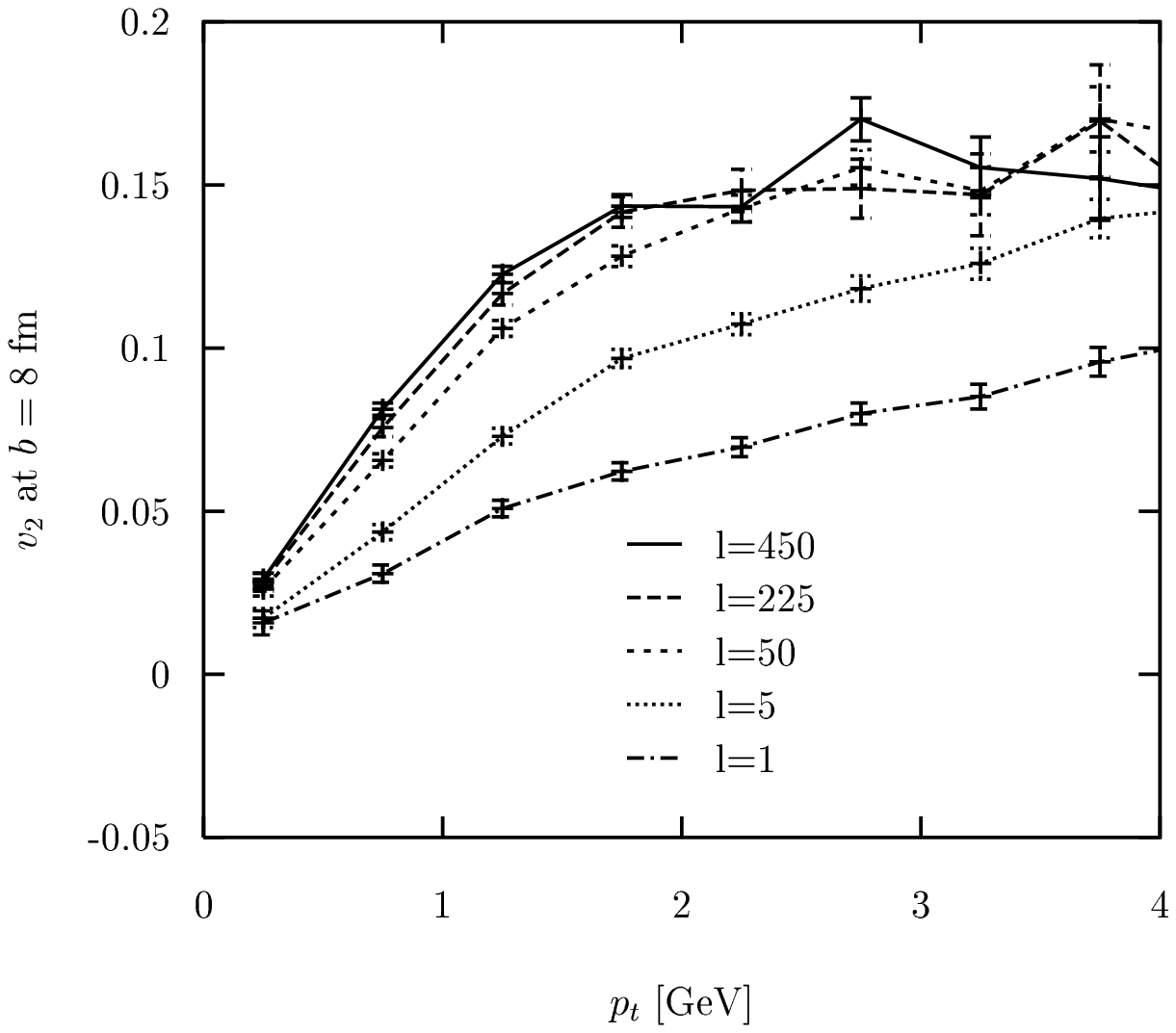}
\hskip 1cm
    \epsfysize 4.5cm
    \epsfbox{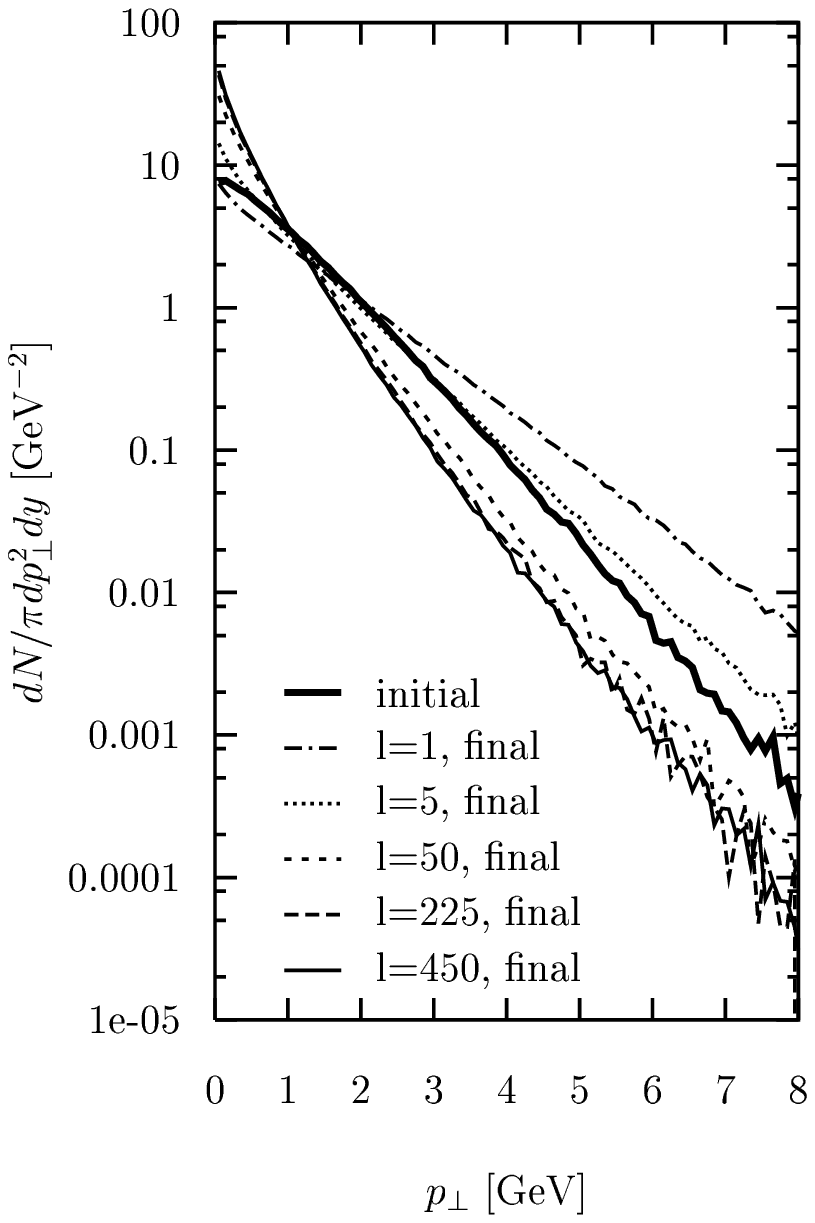}
}
\vspace*{-1cm}
\caption{
\footnotesize
Gluon elliptic flow and $p_\perp$ spectra
for Au+Au at $\sqrt{s}=130A$ GeV with $b=8$ fm
and particle subdivisions $\ell=1$, 5, 50, 225, and 450.
The elastic $gg$ cross section was 100 mb ($\mu = T_0$).
\vspace*{-0.5cm}
}
\label{Figure:3}
\label{Figure:v2_pt_vs_l}
\end{figure}

Fig. \ref{Figure:v2_pt_vs_l} shows that elliptic flow and the
$p_\perp$ spectra are sensitive
to particle subdivision.
To obtain covariant results for these conditions, we needed $\ell > 200$.



\small
\begin{thebibliography}{99}


\bibitem{hydro}
P.~F.~Kolb, P.~Huovinen, U.~Heinz and H.~Heiselberg,
hep-ph/0012137; 
P.~Huovinen, P.~F.~Kolb, U.~Heinz, P.~V.~Ruuskanen and S.~A.~Voloshin,
hep-ph/0101136.

\bibitem{Zhang:1999rs}
B.~Zhang, M.~Gyulassy and C.~M.~Ko,
Phys.\ Lett.\ {\bf B455}, 45 (1999)
[nucl-th/9902016].

\bibitem{molnar_v2}
S.~A.~Bass {\it et al.},
Nucl. Phys. {\bf A661}, 205 (1999) [nucl-th/0005051];
D.~Molnar and M.~Gyulassy,
nucl-th/0102031.


\bibitem{gvw}
X.-N.~Wang,
nucl-th/0009019;
M.~Gyulassy, I.~Vitev and X.-N.~Wang,
nucl-th/0012092.

\bibitem{v2_cascade}
H.~Sorge,
Nucl.\ Phys.\ {\bf A661}, 577 (1999)
[nucl-th/9906051];
M.~Bleicher and H.~Stocker,
hep-ph/0006147.

 
\bibitem{STARv2}
K.~H.~Ackermann {\it et al.}  [STAR Collaboration],
Phys.\ Rev.\ Lett.\ {\bf 86}, 402 (2001)
[nucl-ex/0009011].

\bibitem{nonequil}
D.~Molnar and M.~Gyulassy,
Phys.\ Rev.\ {\bf C 62}, 054907 (2000)
[nucl-th/0005051].
Parton cascade code MPC 1.0.6 used in the present study
can be downloaded from  the OSCAR homepage at
http://www-cunuke.phys.columbia.edu/OSCAR


\bibitem{Gyulassy:1994ew}
M.~Gyulassy and X.~Wang,
Comput.\ Phys.\ Commun.\  {\bf 83}, (1994) 307
[nucl-th/9502021].

\bibitem{Eskola:2000fc}
K.~J.~Eskola, K.~Kajantie, P.~V.~Ruuskanen, and K.~Tuominen,
Nucl. Phys. {\bf B570}, 379 (2000) [hep-ph/9909456].

\bibitem{Yang}
Y.~Pang, RHIC 96 Summer Study, CU-TP-815 preprint (unpublished);
Generic Cascade Program (GCP) documentation available at WWW site
http://www-cunuke.phys.columbia.edu/OSCAR

\bibitem{Zhang:1998ej}
B.~Zhang,
Comput.\ Phys.\ Commun.\  {\bf 109}, 193 (1998)
[nucl-th/9709009].

\bibitem{Zhang:1998tj}
B.~Zhang, M.~Gyulassy, and Y.~Pang,
Phys.\ Rev.\  C {\bf 58}, (1998) 1175
[nucl-th/9801037].








\bibitem{Binnewies:1995ju}
J.~Binnewies, B.~A.~Kniehl and G.~Kramer,
Z.\ Phys.\ C {\bf 65}, 471 (1995)
[hep-ph/9407347].

\end{thebibliography}
\end{document}